\documentclass[aps,prl,showpacs,twocolumn]{revtex4}

\usepackage{graphicx}

\begin{document}

\title{Non-axisymmetric Magnetorotational Instabilities in Cylindrical
Taylor-Couette Flow}

\author{Rainer Hollerbach$^1$, Vijaya Teeluck$^1$ and G\"unther R\"udiger$^2$}
\address{$^1$Department of Applied Mathematics, University of Leeds,
Leeds, LS2 9JT, United Kingdom\\
$^2$Astrophysikalisches Institut Potsdam, An der Sternwarte 16,
D-14482 Potsdam, Germany}

\date{\today}

\begin{abstract}
We study the stability of cylindrical Taylor-Couette flow in the presence
of azimuthal magnetic fields, and show that one obtains non-axisymmetric
magnetorotational instabilities, having azimuthal wavenumber $m=1$.  For
$\Omega_o/\Omega_i$ only slightly greater than the Rayleigh value
$(r_i/r_o)^2$, the critical Reynolds and Hartmann numbers are ${\rm Re}_c
\sim10^3$ and ${\rm Ha}_c\sim10^2$, independent of the magnetic Prandtl
number $\rm Pm$.  These values are sufficiently small that it should be
possible to obtain these instabilities in the PROMISE experimental facility.
\end{abstract}


\maketitle

The magnetorotational instability (MRI) was discovered in 1959 by Velikhov
\cite{V59}, who considered cylindrical Taylor-Couette flow in the presence
of an axial magnetic field, and obtained instabilities in otherwise
hydrodynamically stable flows.  Several decades later, it was recognized
that much the same instability plays a crucial role in the dynamics of
astrophysical accretion disks \cite{BH91}.  This prompted renewed interest
in the MRI in Taylor-Couette flow, and specifically in the possibility of
achieving it experimentally \cite{RZ01,JGK01,RRB04}.  By applying
combined axial and azimuthal magnetic fields, the PROMISE facility
\cite{S06,S07,S08,S09} succeeded in obtaining axisymmetric MRI modes.
We show here that if the magnetic field is taken to be predominantly
azimuthal, one obtains non-axisymmetric modes that should be achievable in
the PROMISE facility.

Taylor-Couette flow, the flow between differentially rotating cylinders,
is one of the most fundamental problems in classical fluid dynamics
\cite{CI94}.  A key result is the so-called Rayleigh criterion, stating that
the flow will be hydrodynamically stable if the angular momentum $\Omega r^2$
increases outward, which occurs whenever the inner and outer cylinder's
rotation rates are adjusted such that $\Omega_o/\Omega_i>(r_i/r_o)^2$.
However, as Velikhov \cite{V59} was the first to show, such flows may
nevertheless be magnetohydrodynamically unstable, provided only that the
angular velocity $\Omega$ decreases outward, that is, $\Omega_o/\Omega_i<1$.
This new instability, now called the magnetorotational instability, has no
analog in the purely hydrodynamic problem, and arises because of the angular
momentum transferred via the magnetic tension in the field lines.

As Balbus and Hawley \cite{BH91} first realized, the MRI may be
critically important in accretion disks, whose Keplerian angular velocity
profiles, $\Omega\sim r^{-3/2}$, are in precisely this regime where the
angular momentum increases outward but the angular velocity decreases.
That is, purely hydrodynamically accretion disks would be stable \cite{JI06},
but magnetohydrodynamically they may be unstable, thereby accounting for the
turbulence and resulting angular momentum transport that is needed to
actually accrete material inward.  See for example \cite{B03} for a recent
review of the MRI in astrophysics.

The recognition of its astrophysical significance led to a resurgence of
interest in the MRI in its original Taylor-Couette context as well, in
particular the possibility of studying it in laboratory experiments.
Following Velikhov, it was originally suggested \cite{RZ01,JGK01} to impose
an axial magnetic field.  However, this `standard' MRI (SMRI) has one very
considerable disadvantage, namely that the rotation rates required to
achieve it are enormous.

The relevant parameter turns out to be not the hydrodynamic Reynolds number
${\rm Re}=\Omega_i r_i^2/\nu$, but rather the magnetic Reynolds number
${\rm Rm}=\Omega_i r_i^2/\eta$, where $\nu$ is the viscosity and $\eta$ the
magnetic diffusivity. The SMRI sets in when ${\rm Rm}\sim10$.  $\rm Re$ is
then given by $\rm Rm/Pm$, where ${\rm Pm}=\nu/\eta$ is the magnetic Prandtl
number, a material property of the fluid.  Typical values are $\sim10^{-5}$
for liquid sodium, and $\sim10^{-6}$ for gallium.  $\rm Re$ must therefore
exceed $10^6$ or even $10^7$, which unfortunately leads to increasingly strong
end-effects \cite{HF04}.  These can perhaps be overcome \cite{SJB09}, but the
SMRI has not been obtained yet.

An alternative approach was suggested by \cite{HR05}, who showed that in a
combined axial and azimuthal magnetic field, the relevant parameter is
$\rm Re$ rather than $\rm Rm$ -- that is, the scaling with $\rm Pm$ is altered
-- and that the resulting `helical' MRI (HMRI) occurs when ${\rm Re}\sim 10^3$,
several orders of magnitude less than what would be required for the SMRI.
This new design was quickly implemented in the PROMISE facility
\cite{S06,S07,S08,S09}, and does indeed yield modes in good agreement with the
theoretical predictions.  Note though that end-effects inevitably play an
important role in this set-up as well, particularly due to the traveling wave
nature of the HMRI.  The implications for the PROMISE results continue to be
debated \cite{Liu09,PG09}.

In this work we start with a purely azimuthal
field.  Velikhov \cite{V59} had already considered this as well, and showed
that it does not yield any axisymmetric instabilities like the SMRI or the
(continuously connected) HMRI.  It can, however, yield non-axisymmetric
instabilities, as \cite{OP96} first demonstrated in an astrophysical context
(where the magnetic fields in accretion disks may indeed be predominantly
azimuthal rather than axial).  The possibility of obtaining non-axisymmetric
instabilities is also particularly exciting, as it would help to circumvent
Cowling's theorem, disallowing purely axisymmetric dynamo action.

In the Taylor-Couette problem considered here, this `azimuthal' MRI (AMRI)
was briefly noted by \cite{RHSE07}, but only in a parameter regime that is
not experimentally accessible.  We show here that for rotation ratios
$\Omega_o/\Omega_i$ only slightly greater than the Rayleigh limit
$(r_i/r_o)^2$, the relevant parameters are sufficiently small that it should
be achievable in the PROMISE facility.

Given the basic state consisting of an azimuthal magnetic field ${\bf B}_0
=B_0(r_i/r){\bf\hat e}_\phi$, imposed by running a current down the central
axis, as well as an angular velocity profile $\Omega(r)$, imposed by
differentially rotating the inner and outer cylinders, we begin by linearizing
the governing equations about it.  The perturbation flow $\bf u$ and field
$\bf b$ may be expressed as
$${\bf u}=             \nabla\times(e{\bf\hat e}_r)
         + \nabla\times\nabla\times(f{\bf\hat e}_r),$$
\vspace{-25pt}
$${\bf b}=             \nabla\times(g{\bf\hat e}_r)
         + \nabla\times\nabla\times(h{\bf\hat e}_r).$$
Taking the $(\phi,z,t)$ dependence to be $\exp(im\phi + ikz + \gamma t)$,
the perturbation equations become
$${\rm Re}\gamma(C_2 e + C_3 f) + C_4 e + C_5 f\qquad\qquad$$
\vspace{-25pt}
$$\qquad\qquad ={\rm Re} E_1 +{\rm Re} F_1 +{\rm Ha}^2 G_1 +{\rm Ha}^2 H_1,$$
\vspace{-20pt}
$${\rm Re}\gamma(C_3 e + C_4 f) + C_5 e + C_6 f\qquad\qquad$$
\vspace{-25pt}
$$\qquad\qquad ={\rm Re} E_2 +{\rm Re} F_2 +{\rm Ha}^2 G_2 +{\rm Ha}^2 H_2,$$
\vspace{-20pt}
$${\rm Rm}\gamma(C_1 g + C_2 h) + C_3 g + C_4 h\qquad\qquad$$
\vspace{-25pt}
$$\qquad\qquad = E_3 + F_3 + {\rm Rm} G_3 + {\rm Rm} H_3,$$
\vspace{-20pt}
$${\rm Rm}\gamma(C_2 g + C_3 h) + C_4 g + C_5 h\qquad\qquad$$
\vspace{-25pt}
$$\qquad\qquad = E_4 + F_4 + {\rm Rm} G_4 + {\rm Rm} H_4.$$
The operators $C_n$ are defined by
$C_n p={\bf\hat e}_r\cdot(\nabla\times)^n(p{\bf\hat e}_r)$,
and work out to be
$$C_1=0,\qquad C_2=\Delta,\qquad C_3=-2mkr^{-2},$$
\vspace{-18pt}
$$C_4=-\Delta\partial_r^2 + (m^2r^{-2}-k^2)(r^{-1}\partial_r-r^{-2})
     + \Delta^2,$$
\vspace{-18pt}
$$C_5=4mk\bigl(r^{-2}\partial_r^2 - r^{-3}\partial_r + (1-m^2)r^{-4}
                 - k^2r^{-2}\bigr),$$
\vspace{-18pt}
$$C_6=\Delta\partial_r^4 - 2(m^2r^{-2}-k^2)r^{-1}\partial_r^3$$
\vspace{-22pt}
$$ + (5m^2r^{-4} - 3k^2r^{-2} - 2\Delta^2)\partial_r^2$$
\vspace{-22pt}
$$ + \bigl(3m^2(2m^2-3)r^{-4} + (4m^2+3)k^2r^{-2} - 2k^4\bigr)r^{-1}
  \partial_r$$
\vspace{-22pt}
$$+m^2(9-10m^2)r^{-6} - 3k^2r^{-4} + 2k^4r^{-2} + \Delta^3,$$
where $\partial_r=\partial/\partial r$, and $\Delta=m^2 r^{-2} + k^2$.
The other quantities are
$$E_1=-im\Delta\Omega e,\qquad E_2=ik\hat\Delta\Omega e,$$
\vspace{-25pt}
$$E_3=0,\qquad E_4=im r^{-2}\Delta e,$$
\vspace{-25pt}
$$F_1=ik(\hat\Delta\Omega + \Delta r\Omega')f,$$
\vspace{-25pt}
$$F_2=-im\Omega(C_4 + 4k^2r^{-2})f-im\Delta(\Omega''+3r^{-1}\Omega')f,$$
\vspace{-25pt}
$$F_3=im r^{-2}\Delta f,\qquad F_4=-i k r^{-2}\hat\Delta f,$$
\vspace{-25pt}
$$G_1=i m r^{-2}\Delta g,\qquad G_2=-i k r^{-2}\hat\Delta g,$$
\vspace{-25pt}
$$G_3=0,\qquad G_4=-im\Delta\Omega g,$$
\vspace{-25pt}
$$H_1=-2i m^2 k r^{-4} h,\qquad H_2=imr^{-2}C_4 h + 4imk^2r^{-4} h,$$
\vspace{-25pt}
$$H_3=-im\Delta\Omega h,\qquad
 H_4=ik(2m^2r^{-2}\Omega-\Delta r\Omega')h,$$
where primes denote $d/dr$, and $\hat\Delta=4m^2 r^{-2} + 2k^2$.
All of these terms are easily derivable using MAPLE, or some other symbolic
algebra package.

Length has been scaled by $r_i$, time by $\Omega_i^{-1}$, $\Omega$ by
$\Omega_i$, $\bf u$ by $\Omega_i r_i$, ${\bf B}_0$ by $B_0$, and $\bf b$ by
${\rm Rm}B_0$.  The two Reynolds numbers $\rm Re$ and $\rm Rm$ are as above;
the Hartmann number ${\rm Ha}=B_0r_i/\sqrt{\rho\mu\eta\nu}$, where $\rho$ is
the fluid's density and $\mu$ the magnetic permeability.  Another parameter
that appears implicitly is the rotation ratio $\hat\mu=\Omega_o/\Omega_i$,
which enters into the details of $\Omega(r)=c_1+c_2/r^2$.  The radius ratio
is fixed at $r_i/r_o=1/2$, as in the PROMISE experiment.

The radial structure of $e$, $f$, $g$ and $h$ was expanded in terms of
Chebyshev polynomials, typically up to $N=30-60$.  These equations and
associated boundary conditions (no slip for $\bf u$, insulating for $\bf b$)
then reduce to a large ($4N\times4N$) matrix eigenvalue problem, with the
eigenvalue being the growth or decay rate $\gamma$ of the given mode.  This
numerical implementation is very different from that of \cite{RHSE07}, in
which the individual components of $\bf u$ and $\bf b$ were used, and
discretized in $r$ by finite differencing.  Both codes yielded identical
results though in every instance where we benchmarked one against the other.

Figure 1 shows the results for $m=1$, the most unstable wavenumber.  At each
point in the $\rm Ha$-$\rm Re$-plane, we repeatedly solve the basic eigenvalue
problem to find the axial wavenumber $k$ that yields the largest ${\rm Re}
(\gamma)$.  We see that if $\hat\mu$ is only slightly greater than the Rayleigh
limit 0.25, values as small as ${\rm Ha}\sim10^2$ and ${\rm Re}\sim10^3$ are
already sufficient to achieve instability.  As $\hat\mu$ is increased,
increasingly large values are required.  Note also that these results are
independent of the Prandtl number; ${\rm Pm}=10^{-5}$, $10^{-6}$, or indeed
even 0 all yield identical results (where we recall that $\rm Pm$ enters the
equations via $\rm Rm=Pm Re$).

The crucial question then is whether ${\rm Ha}\sim10^2$ and ${\rm Re}\sim10^3$
are achievable in the PROMISE facility.  ${\rm Re}\sim10^3$ is certainly
possible; this is precisely the range where the HMRI has already been obtained.
${\rm Ha}=10^2$ is somewhat more challenging, corresponding to a current of 13
kA along the central axis, roughly twice what was required for the HMRI.
Once the latest upgrade is complete though, currents up to 20 kA will be
achievable (F. Stefani, private communication).

Figure 2 quantifies how the critical Reynolds number increases with $\hat\mu$.
That is, we now optimize over $\rm Ha$ as well as $k$, and compute the minimum
value of $\rm Re$ that still allows instability.  The behavior is remarkably
similar to the transition from the HMRI to the SMRI, as shown in Fig.\ 1 of
\cite{HR05}.  In both cases ${\rm Re}_c$ is $\sim10^3$, and independent of
$\rm Pm$ for $\hat\mu$ only slightly greater than the Rayleigh value, but then
increases dramatically, and scales as ${\rm Pm}^{-1}$ once $\hat\mu$ is
sufficiently large.

\begin{figure}
\includegraphics[scale=1.0]{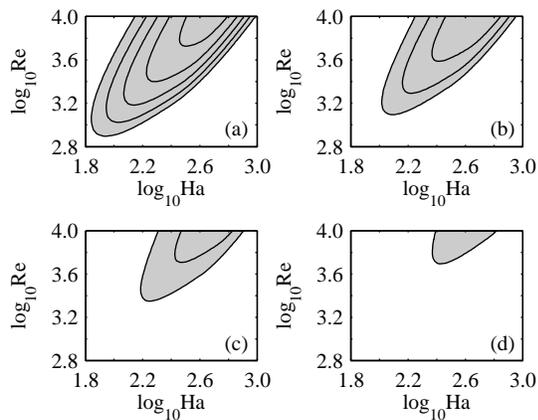}
\caption{The grey-shaded regions show where ${\rm Re}(\gamma)>0$.  The
contour interval is 0.01, indicating that these instabilities grow on the
basic rotational timescale $\Omega_i^{-1}$, but with a somewhat smaller
multiplicative factor than for the SMRI. (a) $\hat\mu=0.25$,
(b) $\hat\mu=0.26$, (c) $\hat\mu=0.27$, (d) $\hat\mu=0.28$.}
\end{figure}

\begin{figure}
\includegraphics[scale=1.0]{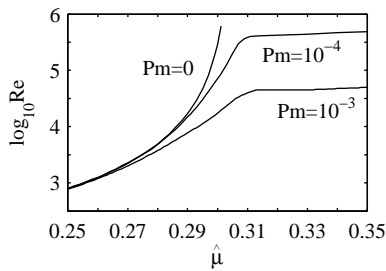}
\caption{${\rm Re}_c$ as a function of $\hat\mu$, optimized over $k$ and
$\rm Ha$.}
\end{figure}

Having obtained this non-axisymmetric instability in a purely azimuthal field,
and demonstrated that it should be achievable in the PROMISE experiment, it is
of further interest to add an axial field again, and investigate at what point
one switches back to the previous axisymmetric HMRI.  We therefore modify the
equations to impose a field of the form ${\bf B}_0=B_0[(r_i/r){\bf\hat e}_\phi
+ \delta{\bf\hat e}_z]$, and explore what happens as $\delta$ is increased
from 0.

At this point we must also consider the handedness of both the basic state and
the resulting instabilities.  For a purely azimuthal field, the basic state has
no handedness, that is, it is invariant to reversing the sign of $z$.  As a
result, instabilities that spiral either to the left (for which $mk>0$) or to
the right (for which $mk<0$), necessarily have exactly the same critical
Reynolds and Hartmann numbers.  For a combined azimuthal and axial field
though, the basic state itself has a handedness \cite{K96,HR05}, so left and
right spiraling instabilities must be considered separately.

Figure 3 shows the results for $\hat\mu=0.26$ and ${\rm Pm}=0$.  An axial field
as weak as $\delta=0.02$ is already enough to induce a clear asymmetry between
the left and right spirals, but both are otherwise still similar to the
$\delta=0$ results from Fig.\ 1(b), included here as the dotted line.  For
$\delta=0.03$ another new feature emerges, the curve labeled 0.  This is
precisely the previous $m=0$, axisymmetric HMRI.  At this value of
$\delta$ the non-axisymmetric modes are still preferred though.  Further
increasing $\delta$, the asymmetry between left and right spirals gradually
becomes greater, and both curves shift upward slightly, indicating that these
modes are suppressed by the addition of an axial field.  In contrast, the HMRI
is strongly excited, so much so that by $\delta=0.05$ it is already the
preferred mode.

\begin{figure}
\includegraphics[scale=1.0]{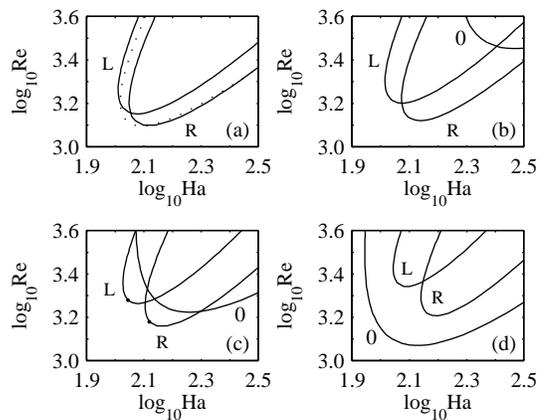}
\caption{The instability curves in the $\rm Ha$-$\rm Re$-plane, as in Fig.\ 1,
but at fixed $\hat\mu=0.26$. (a) $\delta=0.02$, (b) $\delta=0.03$, (c) $\delta
=0.04$, (d) $\delta =0.05$.  Curves labeled L/R denote left/right spiraling
$m=1$ modes, curves labeled 0 the HMRI.  The dotted curve in (a) is at
$\delta=0$, where the L/R curves are identical.  The two dots in (c)
correspond to the solutions shown in Fig.\ 4.}
\end{figure}

We can at least begin to understand why $m=0$ and $1$ behave so differently
by noting that if $m=0$ and $\delta=0$, the instability equation for
the field component $h$ reduces to just free decay, ${\rm Rm}\gamma C_2 h
+C_4h=0$.  However, in the absence of this part of the field
$\nabla\times\nabla\times(h{\bf\hat e}_r)$, there is no radial component to
provide the coupling between different radii that ultimately drives the MRI,
since the other part of the field $\nabla\times(g{\bf\hat e}_r)$ has no
radial component.

This is essentially Velikhov's original proof that a purely azimuthal field
does not yield any axisymmetric instabilities.  See also \cite{HS}, who
extend Velikhov's analysis from ideal to diffusive fluids.  To obtain an
axisymmetric instability, we therefore require $\delta\neq0$.  This couples
$h$ to the other components again, thereby allowing the HMRI to proceed.

The key difference between $m=0$ and $1$ then is that for $m=1$, $h$
is coupled to the other components even if $\delta=0$.  A quick glance at
the instability equations reveals numerous factors of $m$, and hence terms
that drop out for $m=0$ but not for $m\neq0$.  It is this additional
coupling that allows the non-axisymmetric AMRI to exist even in a purely
azimuthal field.

\begin{figure}
\includegraphics[scale=1.0]{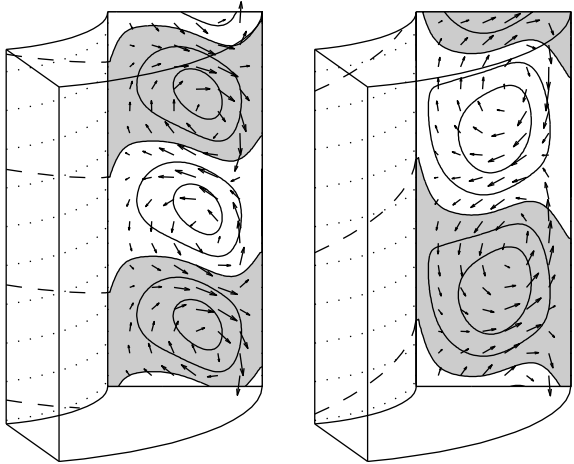}
\caption{The left and right spiral modes, at the two dots indicated in
Fig.\ 3(c).  On the left ${\rm Re}=1910$, ${\rm Ha}=110$, $k=4.2$,
${\rm Im}(\gamma)=-0.24$; on the right ${\rm Re}=1510$, ${\rm Ha}=130$,
$k=-3.0$, ${\rm Im}(\gamma)=-0.27$.  Arrows denote the meridional flow
$(u_r,u_z)$, normalized such that the maximum $(u_r^2 + u_z^2)^{1/2}$ is 1.
Contours show $u_\phi$, with a contour interval of 0.2, grey positive and
white negative.}
\end{figure}

Finally,
Fig.\ 4 shows the spatial structures of the left and right spirals for
$\delta=0.04$.  The arrows indicate the meridional flow $(u_r,u_z)$, and the
contours the azimuthal velocity $u_\phi$.  The dashed lines on the inner
cylinders denote lines of constant phase $m\phi+kz$, so depending on the sign
of $k$ modes spiral either to the left or to the right.  (Alternatively, one
could fix $k$ to be positive, and then consider $m=\pm1$.)  The dotted lines
on the inner cylinders represent a field line of the imposed, slightly
helical field ${\bf B}_0$.  The physical significance of the left/right
asymmetry therefore is that the two modes spiral in the opposite/same
direction as the imposed field.

Future work will consider the nonlinear interactions among these
different modes.  Exactly symmetric left and right spirals in non-magnetic
Taylor-Couette flow already allow a rich variety of possibilities, including
both traveling and standing waves \cite{CI94,LPA03,PLH06}.  It remains to be
seen which of these occurs here for $\delta=0$.  By judiciously adjusting
$\delta$, $\rm Ha$ and $\rm Re$, it should also be possible in this problem
to preferentially select either the left or right modes, or indeed the
axisymmetric HMRI.  The regime $\delta\sim0.04$, where all three modes
have comparable critical Hartmann and Reynolds numbers, is likely to yield
particularly rich dynamics.  Taylor-Couette flows in predominantly azimuthal
magnetic fields of this type clearly deserve further attention, both
experimental and theoretical.

This work was supported by the Science and Technology Facilities Council
under Grant No.\ PP/E001092/1.

\end{document}